\documentclass[twocolumn,showpacs,preprintnumbers,amsmath,amssymb,pre]{revtex4}

\usepackage{graphicx}
\usepackage{dcolumn}
\usepackage{bm}
\usepackage{enumerate}

\begin{document}

\title{Subdiffusion--reaction process with $A\longrightarrow B$ reactions versus subdiffusion--reaction process with $A+B\longrightarrow B$ reactions}

\author{Tadeusz Koszto{\l}owicz}
 \email{tadeusz.kosztolowicz@ujk.edu.pl}
 \affiliation{Institute of Physics, Jan Kochanowski University,\\
         ul. \'Swi\c{e}tokrzyska 15, 25-406 Kielce, Poland}
         
\author{Katarzyna D. Lewandowska}
 \email{kale@gumed.edu.pl}
 \affiliation{Department of Radiological Informatics and Statistics, Medical University of Gda\'nsk,\\ 
 ul. Tuwima 15, 80-210 Gda\'nsk, Poland}

\date{\today}

\begin{abstract}
We consider the subdiffusion--reaction process with reactions of a type $A+B\longrightarrow B$ (in which particles $A$ are assumed to be mobile whereas $B$ --- static) in comparison to the subdiffusion--reaction process with $A\longrightarrow B$ reactions which was studied by I.M. Sokolov, M.G.W. Schmidt, and F. Sagu\'{e}s in \textit{Phys. Rev. E} \textbf{73}, 031102 (2006). In both processes a rule that reactions can only occur between particles which continue to exist is taken into account. Although in both processes a probability of the vanishing of  particle $A$ due to a reaction is independent of both time and space variables (assuming that in the system with the $A+B\longrightarrow B$ reactions, particles $B$ are distributed homogeneously) we show that subdiffusion--reaction equations describing these processes as well as their Greens' functions are qualitatively different. The reason for this difference is as follows. In the case of the former reaction, particles $A$ and $B$ have to meet with some probability before the reaction occurs in contradiction with the case of the latter reaction. For the subdiffusion process with the $A+B\longrightarrow B$ reactions we consider three models which differ in some details concerning a description of the reactions. We base the method considered in this paper on a random walk model in a system with both discrete time and space variables. Then, the system with discrete variables is transformed into a system with both continuous time and space variables. Such a method seems to be convenient in analysing subdiffusion--reaction processes with partially absorbing or partially reflecting walls. The reason is that within this method we can determine Greens' functions without a necessity of solving a fractional differential subdiffusion--reaction equation with boundary conditions at the walls. As an example we use the model to find the Greens' functions for a subdiffusive--reaction system (with the reactions mentioned above), which is bounded by a partially absorbing wall. This example shows how the model can be used to analyze the subdiffusion--reaction process in a system with partially absorbing or reflecting thin membranes. Employing a simple phenomenological model, we also derive equations related to the reaction parameters used in the considered models. 

\end{abstract}

                            
\maketitle


\section{Introduction \label{SecI}}

The process of subdiffusion can occur in media in which particles' movement is strongly hindered due to the internal structure of a medium. Subdiffusion is usually defined as a random walk process in which $\left\langle (\Delta x)^2\right\rangle=2D_\alpha t^\alpha/\Gamma(1+\alpha)$, where $\left\langle (\Delta x)^2\right\rangle$ is a mean square displacement of a random walker, $\alpha$ is a subdiffusion parameter ($0<\alpha<1$), $D_\alpha$ is a subdiffusion coefficient and $\Gamma$ denotes the Gamma function; for normal diffusion there is $\alpha=1$. Subdiffusion is mostly described by means of subdiffusion equations with a fractional time derivative derived from the continuous time random walk formalism \cite{mk}. When the subdiffusion process is extended to a subdiffusion--reaction process then the form and the position of the reaction term within the subdiffusion--reaction equation is not clear. Namely, in paper \cite{sung} the reaction term is located outside the Riemann--Liouville fractional derivative operator whereas in \cite{seki} this term is located under this fractional operator. We should mention here that there are various forms of subdiffusion--reaction equations which are not equivalent (see, for example, \cite{henry,mendez,fedotov1,ah}). The ambiguities concerning the form of the reaction term are related to assumptions concerning the influence of a subdiffusive medium on a reaction kinetic. 

Hereafter, we will consider a subdiffusion process in which a particle can vanish with some probability which does not depend on both time and space variables. Such a process can formally be treated as a subdiffusion--reaction process. Normal diffusion or subdiffusion processes with particles' vanishing occur in biology, for example, in drug absorption via passive diffusion \cite{kimoto} or in a system in which the absorption of particles is weakened by mucus which is treated as a diffusional barrier \cite{larhed}. In order to model such processes, (sub)diffusion--reaction equations with appropriate boundary conditions set at walls bounding the system are needed. However, the choice of boundary conditions is obvious neither for normal diffusion--reaction processes \cite{erban2007} nor for subdiffusive systems in which reactions are absent \cite{kdl2012}. Thus, it is not difficult to realise that the boundary conditions at the walls for a subdiffusion--reaction equation can be set ambiguously. In order to avoid the difficulties with the choice of boundary conditions we propose the following method of modelling subdiffusion--reaction processes. In order to find a subdiffusion--reaction equation and Green's function describing the subdiffusion of a particle $A$ that can also vanish with  some probability,  we will primarily use a random walk model with both discrete time and space variables. Then, the obtained results will be transformed to both continuous time and space variables by means of the formulae presented in this paper. The reason for the choice of such a  methodology is as follows.  Difference equations describing the random walk with both discrete time and space variables can be explicitly solved by means of the generating function method, which means this method can be also applied to study subdiffusion--reaction processes in more complex systems containing various `obstacles' such as partially absorbing or partially reflecting thin membranes. We should mention here that a random walk on a lattice with absorption has been mostly studied by means of the continuous time random walk formalism \cite{henry,mendez,rodriguez,abramson,froemberg,angstmann,ks,felderhof,erban2009,leier}.

In this paper, we will consider two particle's vanishing processes. In the first process a particle $A$ can vanish according to the formula $A\longrightarrow B$. In the second process a particle $A$ has to meet an absorbing centre before the particle's $A$ vanishing takes place. In this case the vanishing process is represented by the reaction $A+B\longrightarrow B$; the absorbing centre is represented by a static particle $B$. Both of the processes can be observed in nature, among other things, in engineering, in biological systems or in ecological systems. The reaction $A\longrightarrow B$ can be observed in filtration processes when particles in any `active state' are moved through a filter. The filter can be a subdiffusive medium in order to give a time to deactivation of particles during its movement inside this medium (see, for example, \cite{cp} and references cited therein). Unimolecular reactions are also an example of such processes. Examples of unimolecular reactions include a radioactive decay, cis-trans isomerization \cite{pvk}, thermal decomposition \cite{sato}, ring opening \cite{bg}, unimolecular nucleophilic substitution reactions \cite{gzg} and racemization \cite{march}. We can also mention here an infected living organism randomly moving in a complex medium; the organism is to die due to the infection. The subdiffusion with $A+B\longrightarrow B$ reaction  can also be observed in a filtration process in which particles react with other particles in a thick filter \cite{hsieh}. The example of the absorbing subdiffusive medium is halloysite, in which absorbing centers are present \cite{la}. As  was mentioned in \cite{ks}, a catalytic quenching of excitations observed in luminescence is also an example of this reaction. The case of an antibiotic treatment is another example of the second reaction. The movement of some bacteria, especially non motile bacteria, can be characterized as a random walk \cite{berg,berg1}. Antibiotics can destroy bacteria in two ways: by killing bacteria or by inhibiting the growth of bacteria \cite{gbfp}. In both cases antibiotics connect in a specific chemical compound occurring in  bacteria \cite{gbfp}.

In the paper \cite{sokolov} Sokolov, Schmidt and Sagu\'{e}s have considered the subdiffusion--reaction process in which a subdiffusive particle $A$ can vanish according to the formula $A\longrightarrow B$ with a constant probability independent of time. Assuming that a probability density $\psi$ of a particle's vanishing reads 
\begin{equation}\label{eq0}
\psi(t)=\gamma{\rm e}^{-\gamma t}, 
\end{equation}
where $\gamma$ is a reaction rate, the authors have shown that the concentration of  particles $A$, $C_A$, is as follows
\begin{equation}\label{eq1}
C_A(x,t)={\rm e}^{-\gamma t}C_{A0}(x,t)\;, 
\end{equation}
where $C_{A0}(x,t)$ is the concentration of particles $A$ in a system in which reactions are switched off, ${\rm e}^{-\gamma t}$ is the probability that a particle continues to exist at time $t$. An important statement presented in the paper cited above is that the concentration is not a solution to the previously mentioned fractional subdiffusion--reaction equations. Therefore, a new equation describing this process is required. The main idea which is the base of the derivation of the new equation (see Eqs. (\ref{eq36}) and (\ref{eq37}) later in this paper) presented in~\cite{sokolov}  is as follows. A reaction can only occur  for particles $A$ which have not vanished before the reaction takes place. A new equation of a somewhat unexpected form has been derived under this assumption. The above considerations lead to the questions: `Should the subdiffusion--reaction equations previously used in many papers be revised?' and `Is this equation still valid when a particle's vanishing is caused by reaction $A+B\longrightarrow B$?' In this paper we will find the answers to these questions.

In every case we will start our considerations with both discrete time and space variables and next we transform them into continuous ones.  Although the model of a subdiffusive system in which mobile particles $A$ can react with static particles $B$ according to the formula $A\rightarrow B$ is presented in the above cited paper, we will consider this process again using a discrete model presented in this paper. This model is based on a random walk model for a system without reactions. The influence of reactions is included in functions describing a particle's waiting time to take its next step. We will pay more attention to subdiffusion processes with $A+B\longrightarrow B$ reactions. To describe this process we will use three models. The first model will be similar to the model which describes the process of subdiffusion with $A\longrightarrow B$ reactions. The previously mentioned  assumption according to which reactions can occur between these particles which continue to exist in the system is included in the model. Within this model we will assume that a reaction can occur at any time (with the reaction probability density (\ref{eq0})) if particle $A$ meets particle $B$ after its jump. `Particles' meeting' here means that particle $A$ will come inside the reaction region of particle $B$ after particle $A$'s jump.  It is assumed that the particles' meeting will occur with some probability $p<1$ (for $p=1$ we obtain results which are equivalent to the case of subdiffusion with $A\longrightarrow B$ reaction). We will show that the introduction of the parameter $p<1$ into the model provides particle concentration which qualitatively differs from (\ref{eq1}). Thus, the models of subdiffusion processes with both types of reactions which have been mentioned above are of a different character. The second model of subdiffusion with $A+B\rightarrow B$ reactions will utilize difference equations  describing random walk, with the particles' vanishing process. This model, in contradiction to the first model,  explicitly includes an absorption probability. We will show that the results obtained within both models mentioned above are similar for $t\gg 1/\gamma$ but the probability distributions obtained within these models differ by the factor $1-p$, which is included into the Green's function obtained for the first model for $t$ less or equal to $1/\gamma$. The third model is equivalent to the second model and assumes that particle $A$ performs an `ordinary' random walk (i.e. just like in a system without  reactions) but the parameters describing the walk depend on the reaction parameter. Additionally, we will find the relation between the parameters occurring in these three models. Finally, we will compare the results obtained within all models mentioned above including the subdiffusion model with the $A\longrightarrow B$ reaction. As an example of extending the considerations beyond the infinite homogeneous system we will derive the probability describing the random walk of a particle in a system bounded by a partially absorbing wall for both of the reactions described above.

The organization of the paper is as follows. In Sec.~\ref{SecIIA} we will present the general procedure which will be used in subsequent considerations.  Starting with the random walk model with a discrete time variable we will show in what way the probability density can be obtained for both continuous time and space variables for the subdiffusion process with reactions. In Sec.~\ref{SecIIB} we will consider subdiffusion in a system without reactions. This section does not have new results but within it we will present some details of the procedure of transferring from discrete variables to continuous ones. In Sec.~\ref{SecIII} we will briefly describe the model presented in \cite{sokolov} and we will show that the procedure presented in Sec.~\ref{SecIIA} provides the results presented in the above cited paper. In Sec.~\ref{SecIV} we will consider subdiffusion with reactions of type $A+B\longrightarrow B$. Starting with the model with both discrete time and space variables, we will derive the subdiffusion--reaction equation for the process under consideration in the system with both continuous  time and space variables. In this section, three varieties of this model will be presented. In Sec.~\ref{SecVA} we will derive the equations related to the reaction parameters used in the models considered in Sec.~\ref{SecIV} by means of the simple phenomenological model. In Sec.~\ref{SecVB} we will compare the results obtained in Secs.~\ref{SecIII} and~\ref{SecIV}. The application of the presented model in describing the subdiffusion for both  types of reactions for a system bounded by a partially absorbing wall will be shown in Sec.~\ref{SecVI}. Final remarks and a discussion of the obtained results will be presented in Sec.~\ref{SecVII}. 

Our considerations concern a three--dimensional system which is homogeneous in the plane perpendicular to the $x$-axis. Thus, later in this paper we treat this system as effectively one--dimensional.

\section{Random walk model of subdiffusion\label{SecII}}

Below, we will show the method of deriving a subdiffusion equation with a fractional time derivative and its fundamental solution (Green's function). We will start our consideration with a random walk in a system in which both time and space variables are discrete. 

\subsection{General model of subdiffusion with reactions\label{SecIIA}}

Supposing $P_n(m;m_0)$ denotes a probability of finding a particle which has just arrived at site $m$ at the $n$--th step; $m_0$ is the initial position of the particle. The random walk is described by the following difference equation 
\begin{equation}\label{geneq}
P_{n+1}(m;m_0)=\sum_{m'} p_{m,m'}P_n(m';m_0)\;,
\end{equation}
where $p_{m,m'}$ is a probability that the particle jumps from the site $m'$, directly to the site $m$. Long jumps can occur with a relatively small probability for subdiffusion or normal diffusion thus, we will take an often applied assumption \cite{hughes,weiss,chandrasekhar} that a jump can only be performed to the neighbouring sites; it is not allowed to stay at the same site at the next moment unless a reflection from the wall occurs. For a random walk without  bias in a homogeneous medium we take the following difference equation
\begin{equation}\label{eq2}
P_{n+1}(m;m_0)=\frac{1}{2}P_n(m+1;m_0)+\frac{1}{2}P_n(m-1;m_0)\;,
\end{equation}
with the initial condition $P_0(m;m_0)=\delta_{m,m_0}$. This equation is usually solved by means of the generating function method \cite{montroll64,barber}. The generating function is defined as 
\begin{equation}\label{eq3}
S(m,z;m_0)=\sum_{n=0}^\infty{z^n P_n(m;m_0)}\;.
\end{equation}
The generating function to Eq.~(\ref{eq2}) reads \cite{barber}
\begin{equation}\label{eq23}
S(m,z;m_0)=\frac{\left[\eta(z)\right]^{|m-m_0|}}{\sqrt{1-z^2}}\;,
\end {equation}
where
\begin{equation}\label{eq24}
\eta(z)=\frac{1-\sqrt{1-z^2}}{z}\;.
\end{equation}

In the subdiffusion model with a continuous time variable $t$, the probability of finding a particle at site $m$ is 
\begin{equation}\label{eq4}
P(m,t;m_0)=\sum_{n=0}^\infty P_n(m;m_0)\Phi_{M, n}(t), 
\end{equation}
where $\Phi_{M,n}(t)$ is the probability that the particle takes $n$ steps over a time interval $(0,t)$ and continues to exist in the system (i.e. the particle has not been absorbed or vanished due to reactions if such processes are present), the index $M$ refers to the model presented later in this paper (for the system without reactions this index will be omitted). The function $\Phi_{M, n}(t)$ depends on the waiting time probability density $\omega_M(t)$ which is needed for the particle to take its next step and continues to exist to time $t$. This function reads 
\begin{equation}\label{n1}
\Phi_{M, n}(t)=\int_0^t U_M(t-t')Q_{M,n}(t')dt', 
\end{equation}
where $U_M(t-t')$ is a probability that the particle has not performed any step over a time interval $(0,t-t')$ (and continues to exists at time $t-t'$) and $Q_{M, n}(t')$ is the probability that the particle performs $n$ steps over this time interval (the last step is performed exactly at time $t'$). The latter function is defined by the following recurrence formula 
\begin{equation}\label{n3}
Q_{M, n}(t')=\int_0^{t'}\omega_M(t'-t'')Q_{M,n-1}(t'')dt'' \;, 
\end{equation}
where $n>1$ and $Q_{M, 1}(t')=\omega_M(t')$. In terms of the Laplace transform, $\mathcal{L}[f(t)]\equiv \hat{f}(s)=\int_0^\infty{{\rm e}^{-st}f(t)dt}$, the function $\Phi_{M, n}(t)$ reads
\begin{equation}\label{eq5}
\hat{\Phi}_{M, n}(s)=\hat{U}_M(s)\hat{\omega}_M^n(s)\;.
\end{equation}
From Eqs.~(\ref{eq3}), (\ref{eq4}) and (\ref{eq5}) we obtain
\begin{equation}\label{new}
\hat{P}(m,s;m_0)=\hat{U}_M(s)S(m,\hat{\omega}_M(s);m_0)\;.
\end{equation}
Equations (\ref{eq23}) and (\ref{new}) provide
\begin{equation}\label{nr1}
\hat{P}(m,s;m_0)=\frac{\hat{U}_M(s)}{\sqrt{1-\hat{\omega}_M
^2(s)}}\;\left[\eta(\hat{\omega}_M(s))\right]^{|m-m_0|}\;.
\end{equation}
From Eqs. (\ref{eq2}), (\ref{eq3}) and (\ref{nr1}) we  obtain the following equation
\begin{eqnarray}\label{eq5b}
[1-\hat{\omega}_M(s)]\hat{P}(m,s;m_0)-\hat{U}_M(s)P(m,0;m_0)\nonumber\\
=\frac{\hat{\omega}_M(s)}{2}\big[\hat{P}(m+1,s;m_0)+\hat{P}(m-1,s;m_0)\nonumber\\
-2\hat{P}(m,s;m_0)\big]\;.
\end{eqnarray}

Supposing $\epsilon$ denotes the distance between discrete sites and supposing 
\begin{equation}\label{iks}
x=\epsilon m\;,\; x_0=\epsilon m_0\;,
\end{equation} 
we transfer variables from discrete to continuous, assuming $\epsilon$ to be small, and use the following relations 
\begin{equation}\label{eq14}
\frac{P(m,t;m_0)}{\epsilon}\approx P(x,t;x_0)\;,
\end{equation}
and
\begin{equation}\label{eq13}
\frac{f(x+\epsilon)+f(x-\epsilon)-2f(x)}{\epsilon^2}\approx \frac{d^2f(x)}{dx^2}\;.
\end{equation}
From Eqs.~(\ref{eq5b})--(\ref{eq13}) we obtain
\begin{eqnarray}\label{b1}
\frac{1-\hat{\omega}_M(s)}{\hat{U}_M(s)}\hat{P}(x,s;x_0)-P(x,0;x_0)\nonumber\\
=\hat{\Psi}_M(s)\frac{\epsilon^2}{2}\frac{\partial^2 \hat{P}(x,t;x_0)}{\partial x^2}\;,
\end{eqnarray}
where 
\begin{equation}\label{b2}
\hat{\Psi}_M(s)=\frac{\hat{\omega}_M(s)}{\hat{U}_M(s)}\;.
\end{equation}
Equation (\ref{b1}) is the base of the derivation of the subdiffusion--reaction equations for the cases considered in this paper.

\subsection{Subdiffusion in a system without reactions\label{SecIIB}}

Considerations presented in this subsection are quite well--known and they are presented here as mathematical preliminaries. In order to obtain the subdiffusion equation one usually considers a mesoscopic model which describes the movement of a single particle and next derives the fractional subdiffusion equation \cite{mk}. In this case the procedure presented in Sec.~\ref{SecIIA} is reduced to the conventional continuous time random walk formalism \cite{mk}. Later we will present a procedure which is a vague contradiction to the previously mentioned one. Namely, we will start our considerations from the fractional equation presented below (see Eq.~(\ref{eq9})) and next, we will find $\hat{\omega}_M(s)$ and $\hat{U}_M(s)$ for which the procedure presented in Sec. \ref{SecIIA} provides this equation (for the case in which reactions are absent we will omit the index $M$ which labels the functions presented in Sec.~\ref{SecIIA}). Considerations which we will present in this section are rather obvious but we will present them here as an illustration of the procedure which will be used in the case of more complicated systems. 

Within the continuous time random walk formalism in which both the waiting time for a particle to take its next step and the steps' length are treated as random variables, subdiffusion can be described by means of the following fractional differential equation with the Riemann--Liouville fractional time derivative. This equation reads (here $0<\alpha<1$) \cite{mk}
\begin{equation}\label{eq9}
\frac{\partial P(x,t;x_0)}{\partial t}=D_\alpha\frac{\partial^{1-\alpha}}{\partial t^{1-\alpha}}\frac{\partial^2 P(x,t;x_0)}{\partial x^2}\;,
\end{equation}
Later in the considerations we assume the following  definition of the subdiffusion coefficient
\begin{equation}\label{eq12}
D_\alpha=\frac{\epsilon^2}{2\tau_\alpha}\;,
\end{equation}
where $\tau_\alpha$ is a parameter given in the units of $1/({\rm sec})^\alpha$, which (together with the dimensionless parameter $\alpha$) characterizes the time distribution $\omega(t)$.
The Riemann--Liouville derivative is defined as being valid for $\delta>0$ (here $k$ is a natural number which fulfils $k-1\leq \delta <k$)
\begin{equation}\label{eq10}
\frac{d^\delta}{dt^\delta}f(t)=\frac{1}{\Gamma(k-\delta)}\frac{d^k}{dt^k}\int_0^t{(t-t')^{k-\delta-1}f(t')dt'}\;.
\end{equation}
The Laplace transform of Eq.~(\ref{eq10}) reads \cite{oldham,podlubny}
\begin{equation}\label{t1}
\mathcal{L}\left[\frac{d^\delta}{d t^\delta}f(t)\right]=s^\delta\hat{f}(s)-\sum_{i=0}^{k-1}s^i f^{(\delta-i-1)}(0)\;,
\end{equation}
where $f^{(\delta-i-1)}(0)$ is the initial value of the derivative of the $(\delta-i-1)$--th order. Since this value is often considered as unknown, the relation (\ref{t1}) is somewhat useless. However, for $0<\delta<1$ and for the case of a bounded function $f$ there is $f^{(\delta)}(0)=0$ therefore, (\ref{t1}) reads for this case
\begin{equation}\label{t3}
\mathcal{L}\left[\frac{d^\delta}{d t^\delta}f(t)\right]=s^\delta\hat{f}(s)\;.
\end{equation}
Thus, the Laplace transform of Eq. (\ref{eq9}) is as follows
\begin{equation}\label{c4}
s\hat{P}(x,s;x_0)-P(x,0;x_0)=s^{1-\alpha} D_\alpha\frac{\partial^2 \hat{P}(x,s;x_0)}{\partial x^2}\;.
\end{equation}
The fundamental solution (Green's function) to Eq. (\ref{eq9}) (i.e. the solution for the initial condition $P(x,0;x_0)=\delta(x-x_0)$) reads in terms of the Laplace transform \cite{mk}
\begin{equation}\label{eq20}
\hat{P}(x,s;x_0)=\frac{1}{2\sqrt{D_\alpha}s^{1-\alpha/2}}{\rm e}^{-\frac{|x-x_0|s^{\alpha/2}}{\sqrt{D_\alpha}}}\;.
\end{equation}
Applying the following formula \cite{koszt}
\begin{eqnarray}\label{eq21}
\lefteqn{\mathcal{L}^{-1}\left[s^\nu {\rm e}^{-as^\beta}\right]\equiv f_{\nu,\beta}(t;a)=}\nonumber\\
&&=\frac{1}{t^{\nu+1}}\sum_{k=0}^\infty{\frac{1}{k!\Gamma(-k\beta-\nu)}\left(-\frac{a}{t^\beta}\right)^k}\;,
\end{eqnarray}
the inverse Laplace transform of Eq. (\ref{eq20}) is
\begin{equation}\label{eq22}
P(x,t;x_0)=\frac{1}{2\sqrt{D_\alpha}}f_{\alpha/2-1,\alpha/2}\left(t;\frac{|x-x_0|}{D_\alpha}\right)\;.
\end{equation}

Now, we are going to determine the functions $\hat{\omega}(s)$ and $\hat{U}(s)$ for the process describing by Eq.~(\ref{eq9}). Comparing Eqs.~(\ref{b1}) and~(\ref{c4}) we obtain
\begin{equation}\label{eq8a}
\hat{U}(s)=\frac{1-\hat{\omega}(s)}{s}\;,
\end{equation}
and
\begin{equation}\label{eq8}
\hat{\Psi}(s)=\frac{s\hat{\omega}(s)}{1-\hat{\omega}(s)}\;.
\end{equation}
Taking into account Eq. (\ref{eq8}), Eq. (\ref{b1}) coincides with Eq. (\ref{c4}) if
\begin{equation}\label{eq16}
\hat{\Psi}(s)=\frac{s^{1-\alpha}}{\tau_\alpha}\;.
\end{equation}
From Eqs. (\ref{eq8}) and (\ref{eq16}) we obtain
\begin{equation}\label{eq17}
\hat{\omega}(s)=\frac{1}{1+\tau_\alpha s^\alpha}\;.
\end{equation}
Equation (\ref{eq17}) provides the Mittag--Leffler time distribution $\omega(t)$ \cite{podlubny}.  
For other distributions, such as the one--sided stable distribution for which $\hat{\omega}(s)={\rm exp}(-\tau_\alpha s^\alpha)$, Eqs. (\ref{eq8}) and (\ref{eq16}) coincide if the Laplace transform of distribution $\omega(t)$ is taken over a limit of small values of $s$ (i.e. assuming that $\tau_\alpha s^\alpha\ll 1$)
which is given by the following general formula \cite{mk} 
\begin{equation}\label{eq19}
\hat{\omega}(s)= 1-\tau_\alpha s^\alpha\;.
\end{equation}
According to the Tauberian theorem and the considerations presented in the Appendix, the limit of small values of $s$ corresponds to the case of a long time limit. 

The discrete model presented above provides not only the subdiffusion differential equation but also directly the probability density (\ref{eq22}). 
From Eqs. (\ref{eq24}) and (\ref{eq19}) we obtain, retaining dominant terms,
\begin{equation}\label{eta1}
\eta\left(\hat{\omega}(s)\right)\approx 1-\sqrt{2\tau_\alpha s^\alpha}\;,
\end{equation}
and from Eqs. (\ref{nr1}), (\ref{eq8a}) and (\ref{eta1}) we get 
\begin{equation}\label{eq25}
\hat{P}(m,s;m_0)= \sqrt{\tau_\alpha}s^{-1+\alpha/2}\left(1-\sqrt{2\tau_\alpha}s^{\alpha/2}\right)^{|m-m_0|}\;.
\end{equation}
Using Eqs. (\ref{iks}), (\ref{eq14}) and  (\ref{eq12})  from Eq. (\ref{eq25}) we obtain Eq. (\ref{eq20}) over a limit of small values of $\epsilon$.

The above calculations  were performed over the limit of small values of $\epsilon$ and were based on Eq. (\ref{eq19}) which is true for $\tau_\alpha s^\alpha\ll 1$. Let us note that the subdiffusion coefficient which together with the subdiffusion parameter $\alpha$ characterizes the subdiffusion process, can be measured experimentally \cite{kdm} and is treated as a constant. Thus, due to Eq. (\ref{eq12}), the limit $\epsilon\rightarrow 0$ results in $\tau_\alpha\rightarrow 0$. Equation (\ref{eq19}) defines function $\omega(t)$ over a long time limit (see Appendix). In the limit $\epsilon\rightarrow 0$, $\omega(s)\rightarrow 1$ and does not define $\omega(t)$ in any way. In order to avoid this ambiguity we can treat $\tau_\alpha$ and, consequently, $\epsilon$ as small but having finite parameters. Taking into account the above circumstances, later in this paper we will treat the parameter $\epsilon$ as finite but small enough in order that Eq.~(\ref{eq20}) would be a proper approximation of Eq.~(\ref{eq25}) taking into account Eqs.~(\ref{iks}), (\ref{eq14}) and~(\ref{eq12}).

\section{Subdiffusion with $A\longrightarrow B$ reactions\label{SecIII}}

Sokolov, Schmidt and Sagu\'{e}s \cite{sokolov} have considered subdiffusion of particles which can vanish according to the formula $A\longrightarrow B$ with a probability independent of concentrations of both particles $A$ and $B$. Assuming that the movement of the particle $A$ is independent of the movement of the other particles, both $A$ and $B$, the general form of the concentration of particles $A$ has been postulated in \cite{sokolov} as Eq.~(\ref{eq1}). This equation has been postulated on the basis of the following idea. It is assumed that the reaction probability density is given by Eq. (\ref{eq0}). The probability that a reaction does not occur over a time interval $(0,t)$ equals ${\rm e}^{-\gamma t}$. In the following process particles which have not vanished before time $t$ can only be involved. The probability $P(x,t;x_0)$  can be obtained for the considered process as a product of the probability that the reaction does not take place and Green's function (\ref{eq22}) and reads
\begin{equation}\label{eqn}
P(x,t;x_0)={\rm e}^{-\gamma t}\frac{1}{2\sqrt{D_\alpha}}f_{\alpha/2-1,\alpha/2}\left(t;\frac{|x-x_0|}{D_\alpha}\right)\;.
\end{equation}

Below we will show that the subdiffusion equation and its solution (\ref{eqn}) can be obtained by means of the procedure presented in Sec.~\ref{SecIIA} (here we will change the index $M$ to $\gamma$). The Laplace transform of Eq. (\ref{eqn}) reads
\begin{equation}\label{eqn1}
\hat{P}(x,s;x_0)=\frac{1}{2\sqrt{D_\alpha}(s+\gamma)^{1-\alpha/2}}{\rm e}^{-\frac{|x-x_0|(s+\gamma)^{\alpha/2}}{\sqrt{D_\alpha}}}\;.
\end{equation}
After calculating the second order derivative of the above function with respect to $x$, and next simple transformations we get the following equation for which function (\ref{eqn1}) is the fundamental solution
\begin{eqnarray}\label{eq30}
& &s\hat{P}(x,s;x_0)-P(x,0;x_0)=\nonumber\\
&=&\frac{\epsilon^2}{2}\hat{\Psi}_\gamma(s)\frac{\partial^2 \hat{P}(x,s;x_0)}{\partial x^2}
-\gamma\hat{P}(x,s;x_0)\;,
\end{eqnarray}
where
\begin{equation}\label{eq35}
\hat{\Psi}_\gamma(s)=\frac{(s+\gamma)^{1-\alpha}}{\tau_\alpha}\;.
\end{equation}
Comparing Eqs. (\ref{b1}) and (\ref{b2}) with (\ref{eq30}) and (\ref{eq35}) we obtain
\begin{equation}\label{eq35a}
\hat{\omega}_\gamma(s)=\hat{\omega}(s+\gamma)\;,\qquad\hat{U}_\gamma=\frac{\tau_\alpha(s+\gamma)^{\alpha-1}}{1+\tau_\alpha(s+\gamma)^\alpha}\;.
\end{equation}
Equation (\ref{eq35a}) provides $\omega_\gamma(t)={\rm e}^{-\gamma t}\omega(t)$ and $U_\gamma(t)={\rm e}^{-\gamma t} U(t)$.

Equations (\ref{eq30}) and (\ref{eq35}) do not provide the fractional differential equation with the Riemann--Liouville fractional derivative for $\gamma\neq 0$. In order to find the equation one takes the following formula into consideration \cite{sokolov}	
\begin{equation}\label{eq34}
\hat{\omega}(s+\gamma)= 1-\tau_\alpha(s+\gamma)^\alpha\;.
\end{equation}
 It has been shown in the paper \cite{sokolov} that the subdiffusion--reaction equation is as follows  using Eqs. (\ref{eq30})--(\ref{eq35})
\begin{equation}\label{eq36}
\frac{\partial P(x,t;x_0)}{\partial t}=D_\alpha\tilde{T}\frac{\partial^2 P(x,t;x_0)}{\partial x^2}-\gamma P(x,t;x_0)\;,
\end{equation}
where the operator $\tilde{T}$ is defined as 
\begin{eqnarray}\label{eq37}
\tilde{T}f(t)=\frac{d}{dt}\int_0^t\frac{{\rm e}^{-\gamma(t-t')}}{(t-t')^{1-\alpha}}f(t')dt'\nonumber\\+\gamma \int_0^t\frac{{\rm e}^{-\gamma(t-t')}}{(t-t')^{1-\alpha}}f(t')dt'\;.
\end{eqnarray}
Equation (\ref{eq36}) can be expressed in the equivalent form which is presented in \cite{henry}, see also Eq. \ref{p9} for $p=1$ later in this paper.

Relation (\ref{eq34}) has a physical meaning if it is treated as an approximation of the Laplace transform of probability density $\hat{\omega}(s)$ over a limit of small values of $\tau_\alpha(s+\gamma)^\alpha$ (unless $\hat{\omega}(s)$ is not given by Eq. (\ref{eq17})). 
Since parameters $\tau_\alpha$ and $\gamma$ are independent, it is not obvious that  $\tau_\alpha(s+\gamma)^\alpha$ has really small values for the small values of $s$. In the case in which $\tau_\alpha \gamma^\alpha$ has large values, Eq.~(\ref{eq36}) along with the operator (\ref{eq37}) are not valid. This problem will be considered elsewhere, however, later in this paper we assume that $\epsilon$ can be chosen to have appropriate small (but nonzero) values in order to assume that --- due to Eq. (\ref{eq12}) --- inequality $\tau_\alpha(s+\gamma)^\alpha\ll 1$ would be fulfilled.

\section{Subdiffusion in a system with $A+B\longrightarrow B$ reactions\label{SecIV}}

In this section we will consider a process in which particle $A$ can vanish due to the reaction of a type $A+B\longrightarrow B$. We additionally assume that the probability of a reaction occurring is independent of both time and space variables. 
The main difference between the reaction under consideration and the reaction considered in the previous section is that the particles $A$ and $B$ have to meet in the reaction region before the reaction takes place.

Below we will consider three models of a random walk with reactions which differ over assumptions concerning  reaction descriptions. In each model our considerations are based on the random walk of a particle $A$ in a system with a discrete space variable. The first model (Model \textit{I}) is based on the assumptions presented in the previous section that a particle $A$ continues to exist at time $t$, but additionally the probability of the particles' meeting is included. The basis of this model is motivated as follows. The discrete system approximates the system with the continuous space variable. Site $m$ in the discrete model corresponds to an interval $(m\epsilon-\epsilon/2,m\epsilon+\epsilon/2)$ in the continuous model. 
Even if we assume that particles $A$ and $B$ are located at the same site in the discrete system, it is not obvious that the particle $A$ is inside the reaction region of the particle $B$. Thus, we will introduce into considerations a probability $p$ that particles $A$ and $B$ meet in the reaction region when they are at the same discrete site. After this meeting the reaction can occur with a probability governed by the probability distribution (\ref{eq0}).  
The second model (Model \textit{II}) is based on the discrete random walk process with a discrete time. Random walk is described by difference equations in which the probability of a reaction occurring is explicitly involved. The probability that the reaction between particles $A$ and $B$ located at the same site at a time between two successive steps of the particle $A$ is given by a single parameter $R$. The jumps between neighbouring sites are governed by $\omega(t)$ which is the same as for processes without reactions.
The third model (Model \textit{III}), which is assumed to be equivalent to Model \textit{II}, brings together the elements occurring in both models. That is to say, it uses the probability $R$ but the considerations are only performed for the case of continuous time using the procedure presented in Sec.\ref{SecIIA}. 

\subsection{Model \textit{I}\label{SecIVA}}

Let us assume that particles $A$ and $B$ can meet with a probability $p$ after a jump of the particle $A$. The reaction can occur over a time interval $(0,t)$ with a probability $(1-{\rm e}^{-\gamma t})p$. Consequently, the probability that the reaction does not take place over this time interval reads $1-(1-{\rm e}^{-\gamma t})p$. The probability density that the particle $A$ makes its jump after time $t$ is the product of the probability that the reaction does not occur and the probability density $\omega(t)$, which gives (here $M=p\gamma$)
\begin{equation}\label{p1}
\omega_{p\gamma}(t)=(1-p)\omega(t)+p{\rm e}^{-\gamma t}\omega(t)\;.
\end{equation}
The Laplace transform of Eq.~(\ref{p1}) reads
\begin{equation}\label{p2}
\hat{\omega}_{p\gamma}(s)=(1-p)\hat{\omega}(s)+p\hat{\omega}(s+\gamma)\;.
\end{equation}
The probability $U_{p\gamma}(t)$ that the particle does not make a jump and continues to exist in the system equals
\begin{equation}\label{p3}
U_{p\gamma}(t)=\left[1-(1-{\rm e}^{-\gamma t})p\right]\left[1-\int_0^t \omega(t')dt'\right]\;.
\end{equation}
The Laplace transform of the above equation reads
\begin{equation}\label{p4}
\hat{U}_{p\gamma}(s)=(1-p)\frac{1-\hat{\omega}(s)}{s}+p\frac{1-\hat{\omega}(s+\gamma)}{s+\gamma}\;.
\end{equation}
From Eqs. (\ref{eq24}), (\ref{eq19}), (\ref{eq34}), (\ref{p2}) and (\ref{p4}) we get
\begin{equation}\label{u1}
\eta\left(\hat{\omega}_{p\gamma}(s)\right)=1-\sqrt{2\tau_\alpha \left[s^\alpha+(s+\gamma)^\alpha\right]}\;.
\end{equation}
and
\begin{equation}\label{u2}
\hat{U}_{p\gamma}(s)=\tau_\alpha\left[s^{\alpha-1}+(s+\gamma)^{\alpha-1}\right]\;.
\end{equation}
Finally, from Eqs. (\ref{nr1}), (\ref{eq12}), (\ref{u1}) and (\ref{u2}) we obtain
\begin{eqnarray}\label{p6}
\hat{P}(x,s;x_0)=\frac{1}{2\sqrt{D_\alpha}}\frac{(1-p)s^{\alpha-1}+p(s+\gamma)^{\alpha-1}}{\sqrt{(1-p)s^{\alpha}+p(s+\gamma)^{\alpha}}}\nonumber\\
\times{\rm e}^{-\frac{|x-x_0|}{\sqrt{D_\alpha}}\sqrt{(1-p)s^{\alpha}+p(s+\gamma)^{\alpha}}}\;.	
\end{eqnarray}
The above function fulfils the following equation
\begin{eqnarray}\label{p7}
& &(1-p)\big[s^{\alpha}\hat{P}(x,s;x_0)-s^{\alpha-1}P(x,0;x_0)\big]\nonumber\\
&+&p\big[(s+\gamma)^{\alpha}\hat{P}(x,s;x_0)-(s+\gamma)^{\alpha-1}P(x,0;x_0)\nonumber\big]\\
&=&D_\alpha\frac{\partial^2 \hat{P}(x,s;x_0)}{\partial x^2}\;.
\end{eqnarray}
In the following we need the Caputo derivative which is defined as being valid for $\alpha>0$: 
\begin{displaymath}
\frac{d^\alpha_C f(t)}{dt^\alpha}=\frac{1}{\Gamma(n-\alpha)}\int_0^t\frac{f^{(n)}(t')dt'}{(t-t')^{\alpha+1-n}}\;,
\end{displaymath} 
where $n$ is a natural number for which $n-1<\alpha<n$.
Taking into account the Laplace transform of the Caputo fractional derivative \cite{podlubny} 
\begin{equation}\label{p8}
\mathcal{L}\left\{\frac{\partial_C^\alpha P(x,t;x_0)}{\partial t^\alpha}\right\}=s^\alpha\hat{P}(x,s;x_0)-s^{\alpha-1}P(x,0;x_0)\;,
\end{equation}
where $0<\alpha<1$, and the following relation
\begin{eqnarray}\label{c5}
\mathcal{L}\left\{ {\rm e}^{-\gamma t}\frac{\partial_C^\alpha}{\partial t^\alpha}{\rm e}^{\gamma t}P(x,t)\right\}
=(s+\gamma)^\alpha\hat{P}(x,s)\nonumber\\
-(s+\gamma)^{\alpha-1}P(x,0)\;,
\end{eqnarray}
we obtain
\begin{eqnarray}\label{p9}
(1-p)\frac{\partial_C^\alpha P(x,t;x_0)}{\partial t^\alpha}+p{\rm e}^{-\gamma t}\frac{\partial_C^{\alpha}}{\partial t^{\alpha}}{\rm e}^{\gamma t} P(x,t;x_0)\nonumber\\
=D_\alpha\frac{\partial^2 P(x,t;x_0)}{\partial x^2}\;.
\end{eqnarray}
Equations (\ref{p6}) and (\ref{p9}) are not really appropriate to practical use. However, for $s\ll\gamma$ (which correspond to $t\gg 1/\gamma$) we get from Eq.~(\ref{p6}) (later in this paper we assume $p<1$)
\begin{eqnarray}\label{p10}
\hat{P}(x,s;x_0)=\frac{1}{2\sqrt{\tilde{D}_\alpha}\mu}s^{\alpha-1}{\rm e}^{-\frac{|x-x_0|\mu}{\sqrt{D_\alpha}}}{\rm e}^{-\frac{|x-x_0|}{\sqrt{D_\alpha}}\frac{(1-p)}{2\mu}s^\alpha},
\end{eqnarray}
where
\begin{equation}\label{g4} 
\tilde{D}_\alpha=D_\alpha/(1-p)\;,\; \mu=\sqrt{p\gamma^\alpha/(1-p)}\;.
\end{equation}
From Eq.~(\ref{p7}) we obtain 
\begin{eqnarray}\label{p11}
& &(1-p)\big[s\hat{P}(x,s;x_0)-P(x,0;x_0)\big]\nonumber\\
&=& s^{1-\alpha}\Bigg[D_\alpha\frac{\partial^2 \hat{P}(x,s;x_0)}{\partial x^2}\nonumber\\
&-&p\gamma^\alpha\big[\hat{P}(x,s;x_0)-(\gamma)^{-1}P(x,0;x_0)\big]\Bigg]\;.
\end{eqnarray}

Equation (\ref{p11}) provides the following equation
\begin{eqnarray}\label{p12}
\frac{\partial P(x,t;x_0)}{\partial t}=\frac{\partial^{1-\alpha}}{\partial t^{1-\alpha}}\Bigg[\tilde{D}_\alpha\frac{\partial^2 P(x,t;x_0)}{\partial x^2}\nonumber\\
-\mu^2 P(x,t;x_0)\Bigg]\;.
\end{eqnarray}
The solution to Eq.~(\ref{p11}) can be obtained from Eqs.~(\ref{eq21}) and~(\ref{p10}) and reads
\begin{eqnarray}\label{p13}
P(x,t;x_0)=\frac{1}{2\mu\sqrt{\tilde{D}_\alpha}}{\rm e}^{-\frac{|x-x_0|\mu}{\sqrt{\tilde{D}_\alpha}}}\nonumber\\
\times f_{\alpha-1,\alpha}\left(t;\frac{|x-x_0|}{2\mu \sqrt{\tilde{D}_\alpha}}\right)\;.
\end{eqnarray}

\subsection{Model \textit{II}\label{SecIVB}}

Considering the subdiffusion--reaction process on a lattice, we assume that if particle $A$ arrives at the site $m$ then it can react with a static particle $B$ located at the same site with the probability $R$. Probability $R$ does not change over time. The generalization of Eq. (\ref{eq2}) to the random walk process with reactions is as follows \cite{abramson}
\begin{eqnarray}\label{eq42}
P_{n+1}(m;m_0)&=&\frac{1}{2}P_n(m+1;m_0)+\frac{1}{2}P_n(m-1;m_0)\nonumber\\
&-&R P_n(m;m_0)\;.
\end{eqnarray}
Using the standard methods presented in \cite{barber,montroll64} we derive the following generating function to Eq. (\ref{eq42})
\begin{equation}\label{eq43}
S(m,z;m_0)=\frac{[\eta_R(z)]^{|m-m_0|}}{\sqrt{(1+zR)^2-z^2}}\;,
\end {equation}
where
\begin{equation}\label{eq44}
\eta_R(z)=\frac{1+zR-\sqrt{(1+zR)^2-z^2}}{z}\;.
\end{equation}
Since the vanishing of a particle caused by a reaction is included in Eq.~(\ref{eq42}) in the term containing parameter $R$, we assume that jumps are governed by $\omega(t)$ (as in the `ordinary' random walk). From Eqs.~(\ref{new}), (\ref{eq42}) and (\ref{eq43}) we obtain
\begin{equation}\label{eq44a}
\hat{P}(m,s;m_0)=\frac{\hat{U}(s)}{\sqrt{[1+R\hat{\omega}(s)]^2-\hat{\omega}^2(s)}}\left[\eta_R(\hat{\omega}(s))\right]^{|m-m_0|}\;.
\end{equation}
From  Eqs. (\ref{eq19}) and (\ref{eq44}) we obtain over a limit of small values of $s$
\begin{equation}\label{eq45}
  \eta_R(\hat{\omega}(s))=\left\{\begin{array}{cc}
      1-\sqrt{2\tau_\alpha s^\alpha}\;,&R=0\;,\\
         \\
      a_R-b_R\tau_\alpha s^\alpha\;,&R\neq 0\;,
    \end{array}
  \right.
\end{equation}
where 
\begin{eqnarray}
a_R&=&1+R-\sqrt{2R+R^2}\;,\label{eq46}\\ 
   \nonumber\\
b_R&=&\frac{(1+R)}{\sqrt{2R+R^2}}-1\;.\label{eq47} 
\end{eqnarray}
In the following we assume that $R\neq 0$. In order to find the probability density for continuous space variables we will conduct the following considerations. Taking into account relations (\ref{eq19}) and (\ref{eq44})--(\ref{eq47}) we obtain
\begin{eqnarray}\label{eq48}
\hat{P}(m,s;m_0)&=&\frac{\tau_\alpha a_R^{|m-m_0|}}{\sqrt{2R+R^2}}s^{\alpha-1}\nonumber\\
&&\times\left(1-\frac{\tau_\alpha s^\alpha}{\sqrt{2R+R^2}}\right)^{|m-m_0|}\;.
\end{eqnarray}
The above equation together with relations (\ref{eq14}), (\ref{eq13}) and (\ref{eq12}) provides
\begin{eqnarray}\label{eq49}
\lefteqn{\hat{P}(x,s;x_0)=\frac{\epsilon s^{\alpha-1}}{2D_\alpha\sqrt{2R+R^2}}}\nonumber\\
&&\times\left(1+R-\sqrt{2R+R^2}\right)^\frac{|x-x_0|}{\epsilon}\nonumber\\
&&\times\left(1-\frac{\epsilon^2 s^\alpha}{2D_\alpha\sqrt{2R+R^2/2}}\right)^{\frac{|x-x_0|}{\epsilon}}\;.
\end{eqnarray}

The only way in order to ensure that function (\ref{eq49}) has non--zero (and finite) values over a limit of small values of  $\epsilon$ is to assume that
\begin{equation}\label{eq50}
\frac{\epsilon}{\sqrt{2R+R^2}}\equiv\frac{1}{\kappa}\equiv const.\;.
\end{equation}
From Eq.~(\ref{eq50}) we obtain
\begin{equation}\label{eq51}
R=\sqrt{1+\epsilon^2\kappa^2}-1\;.
\end{equation}
From Eq.~(\ref{eq51}) we get over a limit of small values of $\epsilon$ ($\epsilon\ll 1/\kappa$)
\begin{equation}\label{eq52}
R=\frac{\epsilon^2\kappa^2}{2}\;,
\end{equation}
and from Eqs. (\ref{eq49}) and (\ref{eq50}) we obtain
\begin{equation}\label{eq53}
\hat{P}(x,s;x_0)=\frac{s^{\alpha-1}}{2D_\alpha\kappa}{\rm e}^{-\kappa|x-x_0|}{\rm e}^{-\frac{|x-x_0|s^\alpha}{2D_\alpha\kappa}}\;.
\end{equation}
From Eqs.~(\ref{eq21}) and (\ref{eq53}) we get
\begin{equation}\label{eq54}
P(x,t;x_0)=\frac{1}{2D_\alpha\kappa}{\rm e}^{-\kappa|x-x_0|}f_{\alpha-1,\alpha}\left(t;\frac{|x-x_0|}{2D_\alpha\kappa}\right)\;.
\end{equation}

Applying the procedure of transforming from both discrete time and space variables to both continuous ones presented in Sec.~\ref{SecIIA} to Eq. (\ref{eq42}) and using Eq. (\ref{eq52}) we obtain 
\begin{eqnarray}\label{eq55}
\frac{\partial P(x,t;x_0)}{\partial t}=D_\alpha\frac{\partial^{1-\alpha}}{\partial t^{1-\alpha}}\Bigg[\frac{\partial^2 P(x,t;x_0)}{\partial x^2}\nonumber\\
-\kappa^2 P(x,t;x_0)\Bigg]\;.
\end{eqnarray}

\subsection {Model \textit{III}\label{SecIVC}}

The assumption that a particle continues to exist in the system at time $t$ is so clear and universal that we should prove that the model presented in Sec. \ref{SecIVB} is not in contradiction with this assumption. With this assumption, the process could be considered as the random walk process in which the probability of a particle's vanishing $R$ is involved in $\omega_R(t)$ (here we change $M$ to $R$). In the model presented in Sec.~\ref{SecIVB} the probability of the vanishing of a particle is included in Eq.~(\ref{eq42}), but the probability density of the waiting time for a particle to take its next step is governed by $\omega(t)$ alone (in which the probability of continuing to exist is not explicitly included). In this section we include the assumption that the particle $A$ which continues to exist will take part in a reaction. It means that the probability that the particle continues to exist is directly included in function $\hat{\omega}_R(s)$ in such a way that Eq.~(\ref{nr1}) provides function~(\ref{eq53}). Therefore, we are looking for a function $\hat{\omega}_R(s)$ depending on the probability $R$ for which the following function
\begin{equation}\label{nr}
\hat{P}(m,s;m_0)=\frac{\hat{U}_R(s)}{\sqrt{1-\hat{\omega}_R^2(s)}}\;\left[\eta(\hat{\omega}_R(s))\right]^{|m-m_0|}\;,
\end{equation}
coincides with (\ref{eq44a}). This coincidence ensures the following equation
\begin{equation}\label{nr2}
\eta_R(\hat{\omega}(s))=\eta(\hat{\omega}_R(s))\;,
\end{equation}
and
\begin{equation}\label{nr3}
\frac{\hat{U}_R(s)}{\sqrt{1-\hat{\omega}_R^2(s)}}=\frac{\hat{U}(s)}{\sqrt{[1+R\hat{\omega}(s)]^2-\hat{\omega}^2(s)}}\;.
\end{equation}
Solution to Eq. (\ref{nr2}) is
\begin{equation}\label{nr4}
\hat{\omega}_R(s)=\frac{\hat{\omega}(s)}{1+R\hat{\omega}(s)}\;,
\end{equation}
and the solution to Eq. (\ref{nr3}) reads taking into account (\ref{nr4})
\begin{equation}\label{nr5}
\hat{U}_R(s)=\hat{U}(s)\frac{1}{1+R\hat{\omega}(s)}\;.
\end{equation}
Taking into account Eq. (\ref{eq19}), from Eqs.~(\ref{nr4}) and~(\ref{nr5})  we obtain  for $\tau_\alpha s^\alpha\ll 1$ 
\begin{equation}\label{nr6}
\hat{\omega}_R(s)= \frac{1}{1+R}\left(1-\tau_{R\alpha}s^\alpha\right)\;,
\end{equation}
and
\begin{equation}\label{nr9}
\hat{U}_R(s)=\tau_{R\alpha}s^{\alpha-1}\;,
\end{equation}
where
\begin{equation}\label{nr7}
\tau_{R\alpha}=\frac{\tau_\alpha}{1+R}\;.
\end{equation}

For the process of subdiffusion (or normal diffusion) controlled reactions, the probability that a reaction takes place is relatively small compared to the probability that a particle makes a jump. In this case we assume that $R\ll 1$, thus $1/(1+R)\approx 1-R$ therefore, we obtain
\begin{equation}\label{nr8}
\hat{\omega}_R(s)=(1-R)\hat{\overline{\omega}}(s)\;,
\end{equation}
and
\begin{equation}\label{nr8a}
\hat{U}_R(s)=\hat{\overline{U}}(s)\;,
\end{equation}
where $\hat{\overline{\omega}}(s)$ and $\hat{\overline{U}}(s)$ are obtained from $\hat{\omega}(s)$ and $\hat{U}(s)$, respectively, after replacing $\tau_\alpha$ by $\tau_{R\alpha}$. 

Functions~(\ref{nr8}) and~(\ref{nr8a}) show that the model presented in Sec.~\ref{SecIVB} is equivalent to the model of the `ordinary' random walk in which the probability density of a particle taking a next step has a form in which the probability of continuing to exist (besides dependence (\ref{nr7})) is included in $\omega_R(t)$ by the factor $1-R$, whereas probability $U_R(t)$ does not include this factor. In this case, we have to accept that the reaction takes place in the considered model just before taking the next step.

In this subsection we have showed in what way model \textit{II} can be presented in terms of functions described in Sec.~\ref{SecIIA}. The equivalence of the models \textit{II} and \textit{III} means that Green's function and the subdiffusion--reaction equation for model \textit{III} are expressed by Eqs.~(\ref{eq54}) and~(\ref{eq55}), respectively. Below, we will show that Eqs.~(\ref{nr}), (\ref{nr4}) and (\ref{nr5}) also provide Eq.~(\ref{eq53}). From Eqs. (\ref{eq24}) and (\ref{nr6})--(\ref{nr7}) we get
\begin{eqnarray}\label{u3}
\eta\left(\hat{\omega}_R(s)\right)=1+R-\sqrt{2R+R^2}
\\-\tau_\alpha s^\alpha \left[\frac{1}{(1+R)\sqrt{2R+R^2}}-1+\frac{\sqrt{2R+R^2}}{1+R}\right]\;.\nonumber
\end{eqnarray}
Equation (\ref{u3}) is transformed to the following form using (\ref{eq51})
\begin{equation}\label{u4}
\eta\left(\hat{\omega}_R(s)\right)=1-\epsilon\left[\kappa+\frac{s^\alpha}{2D_\alpha\kappa}\right]\;.
\end{equation}
From Eqs. (\ref{nr1}), (\ref{iks}), (\ref{eq14}), (\ref{eq12}) and (\ref{u4}) we obtain the function (\ref{eq53}) for small values of $\epsilon$ which the inverse Laplace transform gives Eq.~(\ref{eq54}).

\section{Comparison of the models\label{SecV}}

\subsection{Relation between the reaction parameters \label{SecVA}}

There are various parameters which occur in the models describing the process considered in Sec. \ref{SecIV}. For example, the parameters $\gamma$ and $\kappa$ which characterize reactions are given in various units ($1/$sec and $1/$m, respectively). To find the relation between them let us consider a phenomenological model of subdiffusion with reactions. 
Later in this paper, we utilize one of the simplest models of reactions which consists of the assumption that a reaction can occur when particles $A$ and $B$ meet in the encounter region; the surface of particles $B$ is assumed to be impermeable to particles $A$.
Probability density (\ref{eq0}) characterizes the reaction of type $A+B\longrightarrow B$ in the standard reaction model if particles $A$ and $B$ meet in the encounter region \cite{seki}. 
For the sake of simplicity, we can assume that particles $B$ are spherical and particles $A$ are like points. The encounter region volume for single particle $B$ is $(4/3)\pi[(\rho+b)^3-\rho^3]$, where $\rho$ is the radius of particle $B$ and $b$ being the `thickness' of the encounter region. 
Let us assume for a moment that the model with a continuous space variable is used to describe the random walk with a discrete space variable. The  system is effectively one--dimensional, which corresponds to a three--dimensional system which is homogeneous in a plane perpendicular to the $x$-axis. The three--dimensional system is then divided into cells, each of whose volume is equal to $\Pi\epsilon$, where $\Pi$ is the area of a cell surface which is perpendicular to the $x$-axis. We assume that jumps can be performed between only neighbouring cells which are located at the $x$ axis at interval $(x-\epsilon/2,x+\epsilon/2)$, $x=m\epsilon$, here $m$ being a cell number. During the period between the particle's jumps, it is assumed that the particle does not change its position within the cell. This situation corresponds to the situation in which all objects presented in the above mentioned interval belonging at discrete point $m$, $m=x/\epsilon$, in a discrete system.

The probability that the reaction takes place when the particle $A$ stays within a cell equals $R=p P_R$, where $p$ is the probability that the temporary particle's location is inside the encounter region of one of the particles $B$ and $P_R$ is the probability that the reaction takes place inside this region before the particle $A$ makes its next jump. The volume available to the particle $A$ within the cell equals $\epsilon\Pi-n_B (4/3)\pi \rho^3$, where $n_B$ is the number of particles $B$ inside the cell and the volume of all encounter regions inside the cell equals $n_B (4/3)\pi[(\rho+b)^3-\rho^3]$. The probability that the particle $A$ is located in the encounter region after its jump is the ratio of the volumes and reads 
$p=\mathcal{C}_B \lambda$, where $\mathcal{C}_B=n_B/(\epsilon\Pi)$ denotes the volume concentration of $B$ particles and $\lambda=(4/3)\pi\left[(\rho+b)^3-\rho^3\right]/[1-\mathcal{C}_B(4/3)\pi\rho^3]$. The relationship between one--dimensional $C_B$ and volume concentrations is as follows: $C_B=\Pi\mathcal{C}_B$. If the volume occupied by all particles $B$ located in the cell is significantly smaller compared to the cell's volume (as in the system with a dilute solution of particles $B$) then $\lambda$ can be treated as independent of $\mathcal{C}_B$. For the simplification of the consideration we can assume $\Pi=1$ and later in this paper we will use the quantities defined in the one--dimensional system. If particle $A$ is located within the encounter region, the probability density that a reaction will take place is assumed to be regulated by distribution (\ref{eq0}). The probability that the particle jumps from the reaction region to another cell before the reaction occurring, equals $\int_0^\infty {\rm exp}(-\gamma t)\omega(t)dt=\hat{\omega}(\gamma)$ thus, the probability that the reaction takes place reads $P_R=1-\hat{\omega}(\gamma)$. We define function $F$ as 
\begin{equation}\label{g1}
F(\gamma)=\frac{1-\hat{\omega}(\gamma)}{\tau_\alpha}\;, 
\end{equation}
which is assumed to be analytical (this assumption is motivated by the form of the functions (\ref{eq17}) and (\ref{eq19})).
From Eqs.~(\ref{eq12}), (\ref{eq52}) and (\ref{g1}) we obtain
\begin{equation}\label{eq57}
\kappa^2=C_B\frac{\lambda F(\gamma)}{D_\alpha}\;.
\end{equation}
From Eqs.~(\ref{eq12}), (\ref{eq52}), (\ref{g1}) and (\ref{eq57}) we get
\begin{equation}\label{g2}
R=p\left[1-\hat{\omega}(\gamma)\right]\;.
\end{equation}
Assuming that $\tau_\alpha\gamma^\alpha\ll 1$, from Eqs.~(\ref{eq19}), (\ref{g4}), (\ref{g1}) and (\ref{eq57}) we also obtain
\begin{equation}\label{g3}
\kappa^2=\frac{1-p}{D_\alpha}\mu^2\;.
\end{equation}

\subsection{Differences and similarities of Greens' functions obtained for various models \label{SecVB}}

In Sec. \ref{SecIV} we obtained Greens' functions (\ref{p13}) and (\ref{eq54}) for the models of subdiffusion with $A+B\longrightarrow B$ reactions. The function (\ref{p13}) has been derived under the assumption that a reaction can occur at any time according to the probability distribution (\ref{eq0}) and under the condition that particles $A$ and $B$ meet after a particle $A$ jump. Function (\ref{eq54}) has been obtained within the model based on difference equations. This model is equivalent to the random walk model in which a reaction can take place with probability $R$ just before a particle's jump. Putting (\ref{g3}) into (\ref{p13}) we obtain a function similar to (\ref{eq54}) but with the subdiffusion coefficient controlled by probability $1-p$. Thus, the models provide a function which has similar form to Eq.~(\ref{eq54}). The difference between functions (\ref{p13}) and (\ref{eq54}) can be easily observed over a long time limit in which function (\ref{p13}) reads using Eq. (\ref{g3})
\begin{equation}\label{p14}
P(x,t;x_0)=\frac{1-p}{2D_\alpha \kappa\Gamma(1-\alpha)}{\rm e}^{-\kappa|x-x_0|}\frac{1}{t^\alpha}\;,
\end{equation}
whereas for the second model, from Eqs. (\ref{eq21}) and (\ref{eq54}) we have 
\begin{equation}\label{dlugi2}
P(x,t;x_0)=\frac{1}{2D_\alpha \kappa\Gamma(1-\alpha)}{\rm e}^{-\kappa|x-x_0|}\frac{1}{t^\alpha}\;.
\end{equation}
Function (\ref{p14}) is smaller then function (\ref{dlugi2}) by the factor $1-p$. This factor can be explained taking into account the interpretation of the models. Namely, it has been assumed in the model considered in Sec. \ref{SecIVA}  that the reaction can occur at any time (if a particle $A$ jumps to the reaction region), whereas in the model considered in Sec. \ref{SecIVB} it has been assumed that the reaction can occur just before the particle's next jump.

\begin{figure}[!ht]
\centering
\includegraphics[height=5.7cm]{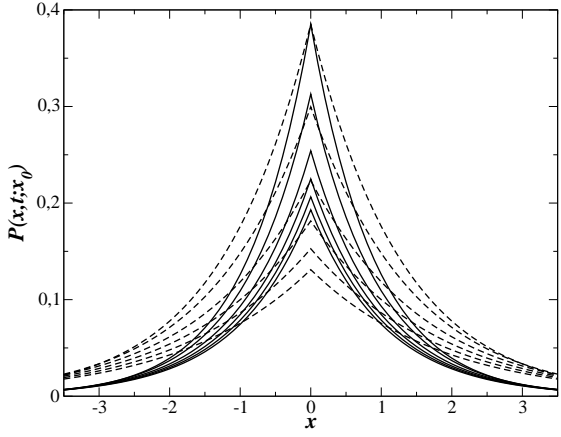}
\caption{The plots of functions (\ref{eqn}) (solid lines) for $\gamma=0.043$ and (\ref{eq54}) (dashed lines) for $\kappa=0.8$ for the following values of time $t\in\{1,2,4,6,8,10\}$; lines which are nearest to the $x$-axis correspond to longer times. In both cases $\alpha=0.3$, $D_\alpha=1.25$ and $x_0=0$ (all quantities are given in arbitrary chosen units).}\label{Fig1}
\end{figure}

The model describing subdiffusion with an $A\longrightarrow B$ reaction provides the equation (\ref{eq36}) and Green's function (\ref{eqn}), which qualitatively differ from the ones obtained within the models which describe the process with an $A+B\longrightarrow B$ reaction.
Below we will compare Greens' functions~(\ref{eqn}) and~(\ref{eq54}). The example plot of functions (\ref{eqn}) and (\ref{eq54}) is presented in Fig.~\ref{Fig1}.
From this figure we can notice that functions obtained for the model of subdiffusion with an $A+B\longrightarrow B$ reaction decrease over time `faster' than the functions obtained for the model with an $A\longrightarrow B$ reaction. This feature is distinctly  manifested in the form of Greens' functions calculated over a long time limit. From Eqs. (\ref{eq21}) and (\ref{eqn}) we obtain
\begin{equation}\label{dlugi1}
P(x,t;x_0)= \frac{1}{2\sqrt{D_\alpha}\Gamma(1-\alpha/2)}{\rm e}^{-\gamma t}\frac{1}{t^{\alpha/2}}\;.
\end{equation}
We can also observe that functions~(\ref{dlugi2}) and~(\ref{dlugi1}) have heavy tails with respect to time. The tail of function~(\ref{dlugi2}) aspires to $0$ faster than the tail of function~(\ref{dlugi1}). Moreover, the values of the above probabilities are suppressed by the exponential terms present in Eqs.~(\ref{dlugi2}) and~(\ref{dlugi1}). For the former model, this is ${\rm e}^{-\kappa|x-x_0|}$, which is independent of time, whereas for the latter model we have ${\rm e}^{-\gamma t}$, which does not depend on a space variable.

\section{Subdiffusion with reactions in a system with partially absorbing wall\label{SecVI}}

\begin{figure}[!ht]
\centering
\includegraphics[height=2cm]{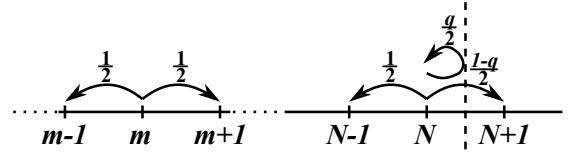}
\caption{Scheme of the system with partially absorbing wall. The numbers above the arrows denote the jumps' probabilities.\label{uklad}}
\end{figure}

Let us consider the subdiffusion of a particle $A$ in a half--space bounded by a partially absorbing wall located between sites $N$ and $N+1$. If the particle tries to jump from the site $N$ to $N+1$ it can be stopped by the wall with the probability $q$ or absorbed by the wall with the probability $1-q$. If the particle passes the wall there is no chance to its returning to the system. This situation is illustrated in Fig. \ref{uklad}.

 This problem can be potentially interesting for various applications mentioned in Sec. \ref{SecI}, for example, if an infected object before its death can leave some area which is surrounded by an imperfect barrier and then an infection gets a chance to spread. Random walk in this system is described by the following difference equations (here $m_0\leq N$)
\begin{equation}\label{ab1}
P_{n+1}(m;m_0)=\frac{1}{2}P_n(m-1;m_0)+\frac{1}{2}P_n(m-1;m_0)\;,
\end{equation}
for $m\leq N-1$ and
\begin{equation}\label{ab2}
P_{n+1}(N;m_0)=\frac{1}{2}P_n(N-1;m_0)+\frac{q}{2}P_n(N;m_0)\;,
\end{equation}
with the initial condition $P_0(m;m_0)=\delta_{m,m_0}$.
The generating function to Eqs. (\ref{ab1}) and (\ref{ab2}) reads
\begin{eqnarray}\label{ab3}
S(m,z;m_0)&=&\frac{[\eta(z)]^{|m-m_0|}}{\sqrt{1-z^2}}
\\&+&\left(\frac{q-\eta(z)}{1-q\eta(z)}\right)\frac{[\eta(z)]^{2N-m-m_0+1}}{\sqrt{1-z^2}}\;.\nonumber
\end{eqnarray}

In the following, we will derive Green's function for the system without reactions. Then, we will find Greens' functions for a subdiffusive system with $A\longrightarrow B$ and $A+B\longrightarrow B$ reactions by means of the method presented in the previous sections.

\subsection{Subdiffusion without reactions\label{SecVIA}}

From Eqs. (\ref{new}) and (\ref{ab3}) we get (as previously, the index $M$ is omitted because this is the case of subdiffusion without reactions)
\begin{eqnarray}\label{ab4}
\hat{P}(m,s;m_0)&=&\frac{\hat{U}(s)[\eta(\hat{\omega}(s))]^{|m-m_0|}}{\sqrt{1-\hat{\omega}^2(s)}}
\\&+&\left(\frac{q-\eta(\hat{\omega}(s))}{1-q\eta(\hat{\omega}(s))}\right)\frac{\hat{U}(s)[\eta(\hat{\omega}(s))]^{2N-m-m_0+1}}{\sqrt{1-\hat{\omega}^2(s)}}\;.\nonumber
\end{eqnarray}

We assume $0<q<1$. From Eqs. (\ref{iks}), (\ref{eq14}), (\ref{eq12}), (\ref{eq8a}), (\ref{eq19}) and using the following approximation $(q-\eta(\hat{\omega}(s)))/(1-q\eta(\hat{\omega}(s)))\approx -1+\sqrt{2\tau_\alpha s^\alpha}(1+q)/(1-q)$ we obtain for small values of $s$
\begin{eqnarray}\label{ab5}
\hat{P}(x,s;x_0)&=&\frac{s^{-1+\alpha/2}}{2\sqrt{D_\alpha}}\left[{\rm e}^{-\frac{|x-x_0|s^{\alpha/2}}{\sqrt{D_\alpha}}}-{\rm e}^{-\frac{(2x_N-x-x_0)s^{\alpha/2}}{\sqrt{D_\alpha}}}\right]\nonumber\\
&+&\left(\epsilon\frac{1+q}{1-q}\right)\frac{s^{\alpha-1}}{2D_\alpha}{\rm e}^{-\frac{(2x_N-x-x_0)s^{\alpha/2}}{\sqrt{D_\alpha}}}\;,
\end{eqnarray}
where $x_N=\epsilon N$.
The last term on the right--hand side of Eq. (\ref{ab4}) vanishes in the limit of small values of $\epsilon$. Then this function appears to be Green's function for the system with a fully absorbing wall. This fact can be explained as follows. The mean number of steps $\left\langle n(t)\right\rangle$ over the time interval $[0,t]$ is expressed by the formula $\left\langle n(t)\right\rangle=\hat{\omega}(s)/[s(1-\hat{\omega}(s))]$ \cite{ks}. Combining this formula with Eqs. (\ref{eq12}) and (\ref{eq19}) we can notice that the jump frequency between neighboured sites goes to infinity when a distance between the sites goes to zero. In this limit the probability that a particle which tries to pass the partially absorbing or partially reflecting wall `infinite times' in every finite time interval, passes the wall in the time interval with probability equals one. Then, the partially absorbing wall behaves as a fully absorbing wall similarly, a partially reflecting wall loses its `reflecting' properties (only a fully reflecting wall or fully absorbing wall do not change their properties when $\epsilon \longrightarrow 0$). In order to keep the permeability properties of the wall over the limit of small values of $\epsilon$, one assumes that the reflecting coefficient $q$ is a function of $\epsilon$. This problem was discussed in \cite{tkoszt2001}, where it was shown that the function $q(\epsilon)$ has an exponential character. Taking into account this result we assume that in a subdiffusive system with a partially absorbing wall there is
\begin{equation}\label{ab6}
q={\rm e}^{-\frac{\epsilon}{\sigma D_\alpha}}\;,
\end{equation}
where $\sigma$ is a `macroscopic' absorbing coefficient of the partially absorbing wall, which can be extracted form experimental data (as well as $D_\alpha$ and $\alpha$). 
Using Eqs. (\ref{eq21}) and (\ref{ab6}), the inverse Laplace transform of Eq. (\ref{ab5}) reads over the limit of small values of $\epsilon$
\begin{eqnarray}\label{ab7}
P(x,t;x_0)&=&\frac{1}{2\sqrt{D_\alpha}}f_{\alpha/2-1,\alpha/2}\left(t;\frac{|x-x_0|}{D_\alpha}\right)\\
&-&\frac{1}{2\sqrt{D_\alpha}}f_{\alpha/2-1,\alpha/2}\left(t;\frac{2x_N-x-x_0}{D_\alpha}\right)\nonumber
\\&+&\sigma f_{\alpha-1,\alpha/2}\left(t;\frac{2x_N-x-x_0}{D_\alpha}\right)\;.\nonumber
\end{eqnarray}

\subsection{Subdiffusion with $A\longrightarrow B$ reaction \label{SecVIB}}

Green's function (here denoted as $P_{A\longrightarrow B}$) for subdiffusion with $A\longrightarrow B$ reaction can be easily obtained by multiplying the Green's function obtained for the system in which a reaction is absent and the probability that a particle continues to exist at time $t$. For the probability density (\ref{eq0}) we get
\begin{equation}\label{ab7a}
P_{A\longrightarrow B}(x,t;x_0)={\rm e}^{-\gamma t}P(x,t;x_0)\;,
\end{equation}
where $P(x,t;x_0)$ is given by Eq. (\ref{ab7}).

\subsection{Subdiffusion with $A+B\longrightarrow B$ reaction \label{SecVIC}}

Green's function for this case can be obtained from Eq.~(\ref{ab4}) by changing $\hat{\omega}(s)\longrightarrow \hat{\omega}_M(s)$ and $\hat{U}(s)\longrightarrow \hat{U}_M(s)$ 
\begin{eqnarray}\label{ab8}
& &\hat{P}(m,s;m_0)=\frac{\hat{U}_M(s)[\eta(\hat{\omega}_M(s))]^{|m-m_0|}}{\sqrt{1-\hat{\omega}_M^2(s)}}\\
&+&\left(\frac{q-\eta(\hat{\omega}_M(s))}{1-q\eta(\hat{\omega}_M(s))}\right)\frac{\hat{U}_M(s)[\eta(\hat{\omega}_M(s))]^{2N-m-m_0+1}}{\sqrt{1-\hat{\omega}_M^2(s)}}\;.\nonumber
\end{eqnarray}
For sufficiently small values of $s$ all models presented in Sec. \ref{SecIV} give the same form of Greens' functions.
Below we use the model presented in Sec. \ref{SecIVC}. Taking into account Eqs. (\ref{eq50}), (\ref{nr6}), (\ref{nr9}), (\ref{nr7}) and (\ref{ab6}), the inverse Laplace transform of Eq. (\ref{ab8}) reads over the limit of small values of $s$ and small values of $\epsilon$ as follows
\begin{eqnarray}\label{ab9}
& & P(x,t;x_0)=\frac{{\rm e}^{-\kappa|x-x_0|}}{2\kappa D_\alpha}f_{\alpha-1,\alpha}\left(t;\frac{|x-x_0|}{2\kappa D_\alpha}\right)\\
&-&\frac{{\rm e}^{-\kappa(2x_N-x-x_0)}}{2\kappa D_\alpha}f_{\alpha-1,\alpha}\left(t;\frac{2x_N-x-x_0}{2\kappa D_\alpha}\right)\nonumber\\
&+&\sigma{\rm e}^{-\kappa(2x_N-x-x_0)}\left[ f_{\alpha-1,\alpha}\left(t;\frac{2x_N-x-x_0}{2\kappa D_\alpha}\right)\right.\nonumber\\
&+&\left.\frac{1}{2\kappa^2 D_\alpha}f_{2\alpha-1,\alpha}\left(t;\frac{2x_N-x-x_0}{2\kappa D_\alpha}\right)\right]\;.\nonumber
\end{eqnarray}

\begin{figure}[!ht]
\centering
\includegraphics[height=5.7cm]{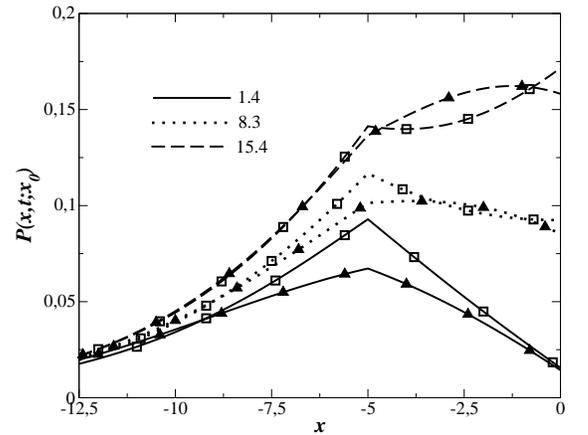}
\caption{The plots of functions (\ref{ab7}) (lines with squares) and (\ref{ab9}) (lines with triangles) for different values of $\sigma$ given in the legend and for $t=100$. The values of the rest of the parameters are as follows $\alpha=0.65$, $D_\alpha=0.75$, $\gamma=0.01$ and $\kappa=0.19$.\label{dwyk8a}}
\end{figure}

\begin{figure}[!ht]
\centering
\includegraphics[height=5.7cm]{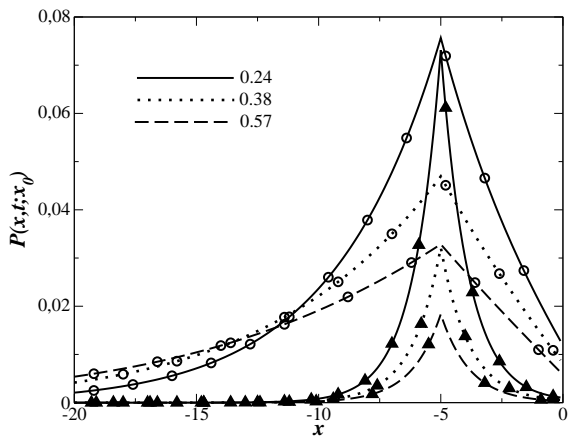}
\caption{The plots of functions (\ref{ab7a}) (lines with circles) and (\ref{ab9}) (lines with triangles) for different values of $\alpha$ given in the legend and for $t=100$. The values of the rest of the parameters are as follows $D_\alpha=2.25$, $\sigma=0.3$, $\gamma=0.008$ and $\kappa=0.83$.\label{dwyk14}}
\end{figure}

\begin{figure}[!ht]
\centering
\includegraphics[height=5.7cm]{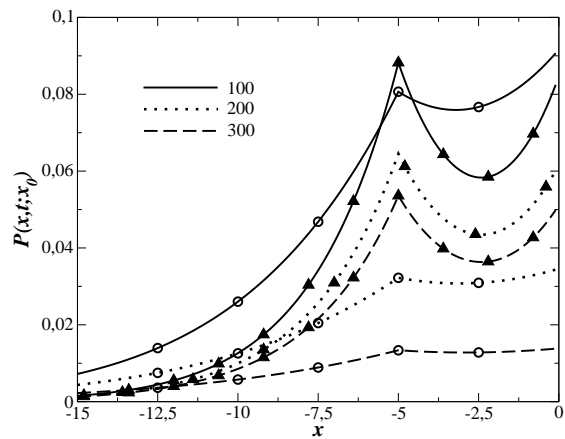}
\caption{The plots of functions (\ref{ab7a}) (lines with circles) and (\ref{ab9}) (lines with triangles) for different values of $t$ given in the legend. The values of the rest of the parameters are as follows $\alpha=0.47$ $D_\alpha=1.25$, $\sigma=7.3$, $\gamma=0.008$ and $\kappa=0.35$.\label{dwyk15}}
\end{figure}

\begin{figure}[!ht]
\centering
\includegraphics[height=5.7cm]{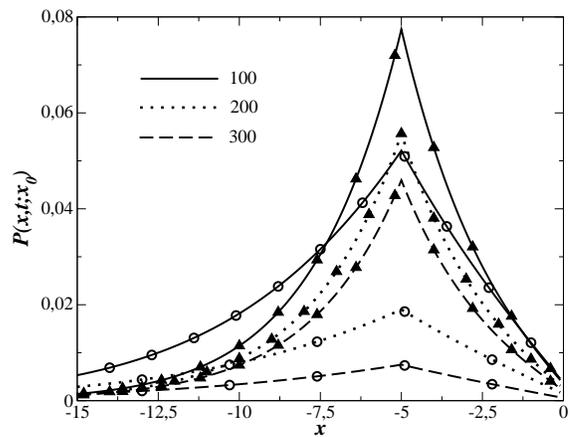}
\caption{The same as in Fig.~\ref{dwyk14} but for $\sigma=0.3$.\label{dwyk16}}
\end{figure}

The example plots of functions (\ref{ab7}), (\ref{ab7a}) and (\ref{ab9}) are presented in Figs. \ref{dwyk8a}--\ref{dwyk16}. For all plots there are $x_0=-5$, $x_N=0$; the values of other parameters are given in each figure separately (all quantities are given in arbitrary chosen units).

In Fig.~\ref{dwyk8a} we present the dependence of functions (\ref{ab7}) and (\ref{ab9}) on the parameter $\sigma$. In the case of function (\ref{ab7}) we can observe an accumulation of a substance near the membrane for higher values of $\sigma$, whereas in the case of function (\ref{ab9}) some amount of substance vanished as a result of reactions and an accumulation is not observed. In Fig.~\ref{dwyk14} we present the dependence of functions (\ref{ab7a}) and (\ref{ab9}) on the parameter $\alpha$. The process occurs faster for higher values of the parameter $\alpha$. Moreover, we can notice that the evolution of the process described by the function (\ref{ab9}) is faster than the second process. In Figs.~\ref{dwyk15} and \ref{dwyk16} we present the dependence of functions (\ref{ab7a}) and (\ref{ab9}) on time. An accumulation of the substance near the membrane can be observed for both functions for a lower value of the parameter $\sigma$, i.e. $0.3$ whereas no accumulation occurs in the case of $\sigma=7.3$.  Summarizing, 'the competition' between the accumulation of particles caused by the selective permeability of the wall and particles' vanishing is observed near the wall. The curves' shapes mainly depend on the relation between parameters $\sigma$ and $\kappa$.

\section{Final remarks\label{SecVII}}

In this paper we have used a random walk model in which both time and space variables are discrete in order to describe subdiffusion processes in which a particle can vanish with some probability. In the first process a particle's vanishing takes place according to the rule $A\longrightarrow B$. The second process concerns subdiffusion with the $A+B\longrightarrow B$ reaction in a system in which static particles $B$ are homogeneously distributed. In both cases a subdiffusive particle $A$ can vanish with a probability which is independent of both time and space variables. 
It has been shown that the process of subdiffusion with $A+B\longrightarrow B$ reactions, which can occur when particle $A$ meets a particle $B$ inside the reaction region, is of a different character in comparison to the process of a particle's vanishing according to $A\longrightarrow B$ reaction. As an example we have considered the subdiffusion with both the above mentioned reactions in a homogeneous system bounded by a partially absorbing wall. 
The example has shown that the discrete model of the subdiffusion--reaction process appears to be a useful tool in modelling subdiffusion--reaction processes in  systems with various `obstacles' such as partially permeable or partially absorbing thin membranes. We have also discussed the properties of the models and we have presented the relationships between parameters occurring in the various models considered in this paper.

The procedure used in this paper which provides the Green's function for a subdiffusion--reaction process in a system with membranes, can be briefly described with the following points.
\begin{enumerate}
	\item We assume difference equations describing the random walk process for the system with fully or partially absorbing or reflecting walls  in which reactions are absent  (such as Eq.~(\ref{geneq}) in Sec.~\ref{SecIIA}) and we find the generating function $S(m,z;m_0)$. We add here that the standard methods of finding this function are presented in \cite{montroll64,barber,hughes,weiss}.
	\item We chose the probability distribution function $\omega_M(t)$ and calculate $U_M(t)$. Next, we substitute the Laplace transforms of these functions and the generating function to Eq. (\ref{new}). Since one expects that the obtained function $\hat{P}(x,s;x_0)$ can be too complex to calculate the inverse Laplace transform analytically, it is recommended  to consider this function in the limit of small values of $s$. Small values of $s$ correspond to large values of $t$ according to the equations presented in the Appendix.
	\item In order to pass from discrete to continuous space variables we use the formulae (\ref{iks}), (\ref{eq14}) and (\ref{eq13}) involving into consideration a definition of subdiffusion coefficient $D_\alpha$ which relates $\epsilon$ to parameters of $\omega(t)$ (we note that Eq. (\ref{eq12}) is an example of such a definition). If a partially absorbing or partially reflecting wall which is characterized by the reflection parameter $q$ is present in the system then we additionally use Eq.~(\ref{ab6}). We add here that an asymmetric wall is characterized by the reflection coefficients $q_1$ and $q_2$ describing a particle's crossing from the right side of the wall to the left side and \textit{vice versa}, respectively. 
\end{enumerate}

Let us note that such a procedure is simpler to use in order to derive Green's function instead of solving subdiffusion--reaction equations (which have a rather complex structure) presented in the previously cited papers  with various boundary conditions at the membranes. Moreover, such boundary conditions appear not to be founded unequivocally (see discussion presented in \cite{kdl2012}). Within the presented method we can proceed inversely. Firstly, we can find Green's function by means of the method presented in this paper, and then we can derive boundary conditions at the walls. We should note that similar methodology was presented by Chandrasekhar in his seminal paper \cite{chandrasekhar} in which boundary conditions at a fully reflecting or fully absorbing wall were derived from Greens' function's for the normal diffusion case.

The derivations of subdiffusion--reaction equations presented in Sec. \ref{SecIV} concern a single particle $A$. Equation~(\ref{eq55}) can be utilized in the derivation of an equation for a large number of particles. Assuming that particles $A$ move independently, their concentration $C_A$ can be calculated according to the formula $C_A(x,t)=\int_{-\infty}^\infty P(x,t;x_0)C_A(x_0,0)dx_0$. Combining this formula with Eqs. (\ref{eq55}) and (\ref{eq57}) we obtain
\begin{eqnarray}\label{eq59}
\frac{\partial C_A(x,t)}{\partial t}=\frac{\partial^{1-\alpha}}{\partial t^{1-\alpha}}\Bigg[D_\alpha\frac{\partial^2 C_A(x,t)}{\partial x^2}\nonumber\\
-k C_A(x,t)C_B\Bigg]\;,
\end{eqnarray}
where $k=\lambda F(\gamma)$. Let us note that Eq. (\ref{eq59}) can be used in a `heuristic derivation' of the subdiffusion--reaction equation for the $A+B\longrightarrow \emptyset(inert)$ reaction (used previously in many papers, see, for example, \cite{yuste,kosztlew}) assuming that the obtained equation is also valid when the concentration of $B$ particles depends on time and space variables, $C_B=C_B(x,t)$.

The universality of the presented method is supported by the fact that various subdiffusion--reaction equations presented in the other papers can be obtained from the equation presented in Sec. \ref{SecIIA}. We can notice that the change of some details concerning the model meaningfully change subdiffusion--reaction equations. For example, in \cite{henry}, various scenarios of the reactions for a subdiffusion system were considered. The main feature of the models studied in \cite{henry} is the moment of the occurring of a reaction when the particle stays at one site. If the particle can instantaneously react with some probability with an other particle just after its jump,  Eq.(43) in \cite{henry} can be obtained with $\hat{U}_M=(1-\tilde{k})\hat{U}(s)$ and $\hat{\omega}_M(s)=(1-\tilde{k})^2\hat{\omega}(s)$ to Eq. (\ref{b1}) presented in this paper, $\tilde{k}$ is a coefficient controlling the vanishing probability of the particle just after its step. For the non-instantaneous annihilation process when a particle reacts at constant per capita rates during the times that it waits before taking its next step, Eq. (60) in \cite{henry} can be obtained from (\ref{b1}) with $\hat{\omega}_M(s)=\hat{\omega}(s+\gamma)$ and $\hat{U}_M(s)=\hat{U}(s+\gamma)$. This situation corresponds for $p=1$ to the model presented in Sec.~\ref{SecIVA}  in this paper which is equivalent to the model presented in \cite{sokolov}. Putting $p=1$ in Eq.~(\ref{p7}) we get an equation equivalent to Eq. (\ref{eq36}) in which operator $\tilde{T}$ is as follows
\begin{equation}\label{eq58}
\tilde{T}f(t)={\rm e}^{-\gamma t}\frac{d^{1-\alpha}}{dt^{1-\alpha}}{\rm e}^{\gamma t}f(t)\;.
\end{equation}
This operator can be expressed by an infinite series of fractional spatial derivatives using Leibniz's formula for the Riemann--Liouville fractional derivative \cite{oldham}. Such an equation has a very complicated structure and in practice it is very hard to analytically treat. For a more general situation in which the particle can be non-instantaneously removed during the waiting times between steps,  Eq. (80) in \cite{henry} can be obtained from (\ref{b1}) taking $\hat{\omega}_M(s)=\mathcal{L}[f(t)\omega(t)]$ and $\hat{U}_M(s)=\mathcal{L}[f(t)U(t)]$, where $f(t)$ is the probability of the surviving of a particle in each step over a time interval $(0,t)$. 

The other assumption presented in \cite{henry} is that the reaction occurs at the same moment as a jump. In this case, the equation derived in \cite{henry} coincides with Eq. (\ref{eq55}) presented in this paper. The interpretation which is presented in Sec.~\ref{SecIVC} is in accordance with the statement presented above according to which the reaction occurs just before the jump. The model {\it I} of subdiffusion with $A+B\longrightarrow B$ reactions provides Eq. (\ref{p9}), which according to the remarks presented above, is not really useful for modelling processes occurring in nature since it is difficult to solve.  We note that Eq.~(\ref{p9}) is different from often using Eq. (\ref{eq55}) (or (\ref{eq59})) which can be obtained within the mean field approximation \cite{ah}. However, as we have shown in this paper, Eq. (\ref{p9}) can be approximated by Eq. (\ref{p11}) for $t\gg 1/\gamma$ which has the same form as Eq. (\ref{eq55}). There arises a question: which equation, (\ref{p9}) or (\ref{p11}), should be taken to describe a subdiffusion--reaction process in which particles have to meet before a reaction? The answer is that it depends on the relation between parameters $\gamma$ and $\tau_\alpha$. The considerations presented in this paper suggest that $\tau_\alpha\longrightarrow 0$ when $\epsilon$ goes to zero due to Eq. (\ref{eq12}). However, the limit $\epsilon\longrightarrow 0$ which can be taken into account passing from discrete to continuous space variables should only be treated as a mathematical trick. In this paper we have treated $\epsilon$ as a small but finite parameter. If we identify $\epsilon$ as a mean length of single particle's jump then $\tau_\alpha$ is defined by Eq. (\ref{eq12}) and takes non-zero values. Let us note that the derivation of a subdiffusion equation within the continuous time random walk formalism \cite{mk} is performed in the limit of small values of parameter $s$, $\tau_\alpha s^\alpha\ll 1$, assuming that $\tau_\alpha$ is finite \cite{mk}, which is in accordance with the just mentioned remarks. Thus, it may be the case that $1/\gamma$ is comparable to $\tau_\alpha^{1/\alpha}$ or smaller. Then Eq. (\ref{p11}) can be used instead of Eq.~(\ref{p9}), for times for which the continuous time random walks formalism works. In this case subdiffusion-reaction equations derived within the models {\it I},  {\it II} and {\it III} have the same form in spite of the fact that the assumptions concerning the moment of the occurrence of a reaction during a particle's staying at one site are different for each of the models. Because the models {\it I},  {\it II} and {\it III} differ on these assumptions, we conclude that the form of the equation does not depend on the moment of reaction occurring when $1/\gamma \ll \tau_\alpha^{1/\alpha}$. 

The considerations presented in this paper justify that subdiffusion--reactions equations are equations of a different kind for the considered processes and that an unexpected form of the subdiffusion--reaction equation for $A\longrightarrow B$ derived in \cite{sokolov} does not prohibit the subdiffusion--reaction process in which particles have to meet with the probability $p<1$ in order to react can be described by means of such an equation as~(\ref{eq59}).

It is obvious that the usefulness of a subdiffusion--reaction equation should be verified by comparison of experimental and theoretical results. We note that various assumptions concerning a moment of reaction occurring is beyond experimental verification. There are theoretical functions provided by Eq. (\ref{eq59}) which coincide well with the experimental results. The example is the time evolution of carious lesion progress \cite{kosztlew2}. As far as we know, an equation in the form of~(\ref{p9}) (for $p=1$ as well as $p<1$) has not been experimentally verified yet. However, it is possible that this equation  (or other subdiffusion--reaction equations presented in other papers) can be applied to model subdiffusion processes with reactions, especially when parameter $\gamma$ is relatively small.
Since experimental investigations concerning subdiffusion are frequently conducted for membrane systems \cite{kdm} and references cited therein, we therefore, hope our model, which can be utilized for the membrane systems, will facilitate a possible experimental verification of models.

\appendix

\section{Laplace transform for small parameter}

We will show an approximation of the Laplace transform for small values of the parameter $s$ which is based on Eq. (\ref{eq21}).
Substituting ${\rm e}^{-as^\beta}=\sum_{j=0}^\infty (-a)^j s^{j\beta}/j!$ to Eq.~(\ref{eq21}) and taking $\nu=0$ we get
\begin{eqnarray}\label{ap1}
\mathcal{L}^{-1}\left[1-as^\beta +\frac{a^2}{2}s^{2\beta}+\ldots\right]\\ 
= -\frac{a}{\Gamma(-\beta)t^{\beta+1}}+\frac{a^2}{2\Gamma(-2\beta)t^{2\beta+1}}+\ldots\;.\nonumber
\end{eqnarray}
The more formally proof of Eq. (\ref{ap1}) was presented in \cite{krylov}. In the following, we assume that $0<\beta\leq 1$.
The condition 
\begin{equation}\label{ap2}
as^\beta\ll 1
\end{equation}
implies that every term on the left--hand side of Eq.~(\ref{ap1}) containing $s^{j\beta}$ is significantly smaller than a term containing $s^{(j+1)\beta}$. Thus, one can approximate the series by the two first terms. Then, the series located on the right--hand side of this equation can be approximated by the first term alone. This approximation is acceptable if 
\begin{equation}\label{ap3}
\left|\frac{a}{\Gamma(-\beta)t^{\beta+1}}\right|\gg \left|\frac{a^2}{2\Gamma(-2\beta)t^{2\beta+1}}\right|\;.
\end{equation}
Using the formula $\Gamma(2z)/\Gamma(z)=2^{2z-1/2}\Gamma(z+1/2)/\sqrt{2\pi}$, from (\ref{ap3}) we get
\begin{equation}\label{ap4}
t\gg \Lambda_\beta a^{1/\beta}\;,
\end{equation}
where
\begin{equation}\label{ap5}
\Lambda_\beta= \left[\frac{\sqrt{\pi}2^{2\beta}}{|\Gamma(-\beta+1/2)|}\right]^{1/\beta}\;.
\end{equation}
Numerical calculations have shown that $\Lambda_\beta$ decreases the function of $\beta$ for $\beta\in (0,1/2)$ and increases the function for $\beta\in (1/2,1]$. There is $\Lambda_\beta\longrightarrow 1$ when $\beta\longrightarrow 0$, $\Lambda_{1/2}=0$ and $\Lambda_1=2$. The example of the other values are $\Lambda_{0.745}\approx 1.00$, $\Lambda_{0.8}=1.31$, and $\Lambda_{0.9}=1.75$. Thus, it seems to be appropriate to approximate the condition (\ref{ap4}) by the following
\begin{equation}\label{ap6}
t\gg a^{1/\beta}\;,
\end{equation}
which can be alternatively replaced by $t\gg 2a^{1/\beta}$ if $\beta$ is close to $1$. Under these conditions we get from (\ref{ap1})  $\mathcal{L}^{-1}[1-as^\beta]\approx -a/[\Gamma(-\beta)t^{\beta+1}]$ (the expression $1-as^\beta$ is treated here as an approximation of the Laplace transform of a function, it is not possible to calculate an inverse Laplace transform of this expression term by term). Taking into account Euler's reflection formula $\Gamma(1-z)\Gamma(z)=\pi/{\rm sin}(\pi z)$, we obtain $\mathcal{L}^{-1}[1-as^\beta]\approx a\;{\rm sin}(\pi\beta)\Gamma(1+\beta)/(\pi t^{\beta+1})$.

\end{document}